\documentstyle[12pt]{article}

\begin{document}

\renewcommand{\thefootnote}{\fnsymbol{footnote}}

\hsize37truepc\vsize61truepc
\hoffset=-.5truein\voffset=-0.8truein
\setlength{\baselineskip}{17pt plus 1pt minus 1pt}
\setlength{\textheight}{22.5cm}

\def\diag{{\rm diag}}
\def\I{{\rm i}}
\def\tr{{\rm tr}}
\def\boldomega{\mbox{\boldmath $\omega$}}

\begin{titlepage}

\noindent December, 1994 \hfill{MRR 080-94}\\
\mbox{ } \hfill{hep-th/9502039}
\vskip 1.6in
\begin{center}
{\Large {\bf Diagonal $K$-matrices and transfer matrix }}\\[8pt]
{\Large {\bf   eigenspectra associated with the $G^{(1)}_2$ $R$-matrix}}
\end{center}

\normalsize
\vskip .4in

\begin{center}
C. M. Yung  \hspace{3pt}
and \hspace{3pt} M. T. Batchelor
\par \vskip .1in \noindent
{\it Department of Mathematics, School of Mathematical Sciences}\\
{\it Australian National University, Canberra ACT 0200, Australia}
\end{center}
\par \vskip .3in

\begin{center}
{\Large {\bf Abstract}}\\
\end{center}

We find all the diagonal $K$-matrices for the $R$-matrix associated with
the minimal representation of
the exceptional affine algebra $G^{(1)}_2$. The corresponding transfer
matrices are diagonalized with a variation of the analytic Bethe ansatz.
We find many similarities with the case of the Izergin-Korepin $R$-matrix
associated with the affine algebra $A^{(2)}_2$.

\end{titlepage}

\section{Introduction}
A systematic method, which extends the Quantum Inverse Scattering Method,
has been developed by Sklyanin \cite{Sklyanin88} and others
\cite{Mezincescu91a} to construct integrable
models with open boundaries. The procedure begins with an $R$-matrix $R(u)$,
which by definition is a solution of the Yang-Baxter equation (YBE). As is
well-known,  given such a solution it is possible to construct a commuting
transfer matrix which defines an integrable vertex model with periodic
boundaries.  For open boundary versions of this
model one requires solutions of a boundary version of the YBE,
or `reflection-factorization equation'.  Given $R(u)$
and a solution $K^-(u)$ of this equation, known as a $K$-matrix, a
commuting transfer matrix $t(u,\boldomega)$ can be defined -- under certain
fairly general technical assumptions on $R(u)$. The parameters
$\boldomega=(\omega_1,\ldots,\omega_N)$
in the transfer matrix are ``inhomogeneities''; setting them to
zero and taking the logarithmic derivative yields an integrable open
spin chain \cite{Sklyanin88},
whereas setting them to alternate as $\omega_i=(-)^i u$ yields
an integrable vertex model with open boundaries \cite{Destri92,Yung94}.

The boundary YBE has been investigated for various $R$-matrices. In
particular, it is known \cite{Mezincescu91b} that $K^-(u)=1$ is a solution
for all the $R$-matrices associated with the vector representations of the
non-exceptional affine Lie algebras \cite{Bazhanov85,Jimbo86} in the
homogeneous gauge, except for the ones associated with $D^{(2)}_{n+1}$.
For the $A^{(1)}_n$ family, all the diagonal $K$-matrices are known
\cite{deVega93} and the corresponding transfer matrices have been
diagonalized \cite{deVega94a}. All diagonal $K$-matrices and transfer matrix
eigenspectra are also known for the three-state models associated with
$A^{(2)}_2$ (the Izergin-Korepin model)
\cite{Mezincescu91b,Mezincescu92b,Yung94} and
the spin-1 representation of $A^{(1)}_1$ (the Zamolodchikov-Fateev
model) \cite{Mezincescu90}. Very recently, the transfer matrices
corresponding to the $K^-(u)=1$ solution for $A^{(2)}_{2n}$ have been
diagonalized \cite{Artz94}.

In this paper we study the $R$-matrix \cite{Kuniba90} associated with the
minimal representation of $G^{(1)}_2$, being the smallest $R$-matrix
associated with an exceptional affine algebra. We show that the $R$-matrix
satisfies all the requirements given in \cite{Mezincescu91a} for the
construction of a commuting Sklyanin transfer matrix. In particular,
$K^-(u)=1$ is a solution to the boundary YBE. We find, in addition,
that there are only two other inequivalent diagonal solutions. For all
three $K$-matrices we obtain the corresponding transfer matrix eigenspectra,
using a modification of the analytic Bethe ansatz very similar to that
employed in \cite{Artz94}. We find that the Bethe ansatz equations obtained
are very similar to those for the Izergin-Korepin model
\cite{Mezincescu92b,Yung94}, which have recently \cite{Batchelor94} found
applications to surface critical phenomena in self-avoiding walks.

\section{The $G^{(1)}_2$ $R$-matrix and related $K$-matrices}
The $G^{(1)}_2$ $R$-matrix $R(u)$ acts in the tensor product of the
(seven-dimensional) representation $V_{\Lambda_2}$ of $U_q(G_2)$ with
itself. In terms of the projectors ${\cal P}_{\Lambda}$ onto the irreducible
representations $V_{\Lambda}$ occurring in the $U_q(G_2)$-module decomposition
$V_{\Lambda_2} \otimes V_{\Lambda_2} = V_0 \oplus V_{\Lambda_1} \oplus
V_{\Lambda_2} \oplus V_{2\Lambda_2}$,
we have the expression \cite{Kuniba90}
\begin{equation}
\check{R}(u) = \sum_{\Lambda=2\Lambda_2,\Lambda_1,\Lambda_2,0}
\rho_{\Lambda}(u) {\cal P}_{\Lambda}
\end{equation}
for $\check{R}(u)=P R(u)$,
with $P$ being the permutation operator and the functions
$\rho_{\Lambda}(u)$ given by
\begin{displaymath}
\rho_{\Lambda}(u) = \left\{\begin{array}{cc}
   {[}1+u{]}{[}4+u{]}{[}6+u{]}  \hspace{10pt} & \Lambda=2\Lambda_2\\
   {[}1-u{]}{[}4+u{]}{[}6+u{]}  \hspace{10pt} & \Lambda=\Lambda_1\\
   {[}1+u{]}{[}4-u{]}{[}6+u{]}  \hspace{10pt} & \Lambda=\Lambda_2\\
   {[}1-u{]}{[}4+u{]}{[}6-u{]}  \hspace{10pt} & \Lambda=0 \end{array}\right.,
\end{displaymath}
where ${[}x{]}\equiv (q^x-q^{-x})(q-q^{-1})^{-1}$. Explicit expressions for the
projectors ${\cal P}_{\Lambda}$ in terms of $q$-Wigner coefficients
can also be found in \cite{Kuniba90}.
The $R$-matrix defines an integrable seven-state 175-vertex model with
periodic boundaries. There also exists a related integrable RSOS model
\cite{Kuniba91}, with an elliptic generalization.

For the transfer matrix
$t_{{\rm P}}(u,\boldomega) = \tr_a T_a (u,\boldomega)$
with periodic boundary conditions (the monodromy matrix $T_a(u,\boldomega)$ is
defined in (\ref{eqn:monod}))
the eigenvalue expression $\Lambda_{{\rm P}} (u,\boldomega)$
was conjectured in \cite{Reshetikhin87} on the basis of Dynkin diagram
considerations, before the explicit form of the $R$-matrix was known. It has
since been confirmed using the analytic Bethe ansatz in a recent study
\cite{Suzuki94} in which the fused models were also considered. We present it
here for later comparison.  Introduce the function $f(u)=\prod_{j=1}^N
[u+\omega_j]$, through which are
defined
\begin{eqnarray}
\phi_{-3}(u,\boldomega) & = & f(1+u) \; f(4+u) \; f(6+u), \nonumber\\
\phi_{-2}(u,\boldomega)=\phi_{-1}(u,\boldomega) & = &
      f(u) \; f(4+u) \; f(6+u),\nonumber \\
\phi_{0}(u,\boldomega) & = & f(u) \; f(3+u) \; f(6+u),\label{eqn:rel}\\
\phi_{1}(u,\boldomega) = \phi_{2}(u,\boldomega) & = &
   f(u) \; f(2+u) \; f(6+u),\nonumber\\
\phi_{3}(u,\boldomega) & = & f(u) \; f(2+u) \; f(5+u).\nonumber
\end{eqnarray}
Define also the functions
$d^{(i)}(u) = \prod_{j=1}^{N_i} [u-\I u_j^{(i)}]$ for $i=1,2$.
The eigenvalue expression is then given by
\begin{equation}
\Lambda_{{\rm P}}(u,\boldomega)= \sum_{j=-3}^{3} \phi_j(u,\boldomega) a_j(u),
\end{equation}
where we have defined
\begin{eqnarray}
a_{-3}(u) & = & \frac{d^{(2)}(u-1/2)}{d^{(2)}(u+1/2)}, \nonumber\\
a_{-2}(u) & = & \frac{d^{(1)}(u-1) \; d^{(2)}(u+3/2)}
    {d^{(1)}(u+2) \; d^{(2)}(u+1/2)}, \hspace{15pt}
a_{-1}(u)  =  \frac{d^{(1)}(u+5) \; d^{(2)}(u+3/2)}
    {d^{(1)}(u+2) \; d^{(2)}(u+7/2)}, \nonumber\\
a_{0}(u) & = & \frac{d^{(2)}(u+9/2) \; d^{(2)}(u+3/2)}
    {d^{(2)}(u+5/2) \; d^{(2)}(u+7/2)}, \nonumber\\
a_{1}(u) & = & \frac{d^{(1)}(u+1) \; d^{(2)}(u+9/2)}
    {d^{(1)}(u+4) \; d^{(2)}(u+5/2)}, \hspace{15pt}
a_{2}(u)  =  \frac{d^{(1)}(u+7) \; d^{(2)}(u+9/2)}
    {d^{(1)}(u+4) \; d^{(2)}(u+11/2)},\nonumber\\
a_{3}(u) & = & \frac{d^{(2)}(u+13/2)}{d^{(2)}(u+11/2)}.\label{eqn:dD}
\end{eqnarray}

The following are properties \cite{Kuniba90} of the $R$-matrix:
\begin{eqnarray}
{\rm commutativity} \hspace{10pt} & : & \hspace{10pt}
     [\check{R}_{12}(u),\check{R}_{12}(v)]=0, \label{eqn:comm}\\
{\rm unitarity} \hspace{10pt} & : & \hspace{10pt}
     R_{12}(u) R_{12}^{t_1 t_2}(-u) = \xi(u), \label{eqn:unit}\\
{\rm regularity} \hspace{10pt} & : & \hspace{10pt}
     R_{12}(0) = \xi(0)^{1/2} P_{12},\\
{\rm PT-symmetry} \hspace{10pt} & : & \hspace{10pt}
     R_{21}(u) \equiv P_{12} R_{12}(u) P_{12} = R_{12}^{t_1 t_2}(u),\\
{\rm crossing-symmetry} \hspace{10pt} & : & \hspace{10pt}
     R_{12}(u) = - \stackrel{1}{V} R_{12}^{t_2}(-u-\rho) \stackrel{1}{V}.
\label{eqn:cross}
\end{eqnarray}
In the above relations, we have $\xi(u)=
{[}1+u{]}{[}4+u{]}{[}6+u{]}{[}1-u{]}{[}4-u{]}{[}6-u{]}$; the
crossing-parameter is $\rho=6$, and the
crossing-matrix $V=F(\sigma g)^2$ satisfying
$V^2=1$. We also use the notation $\stackrel{1}{A}$ to denote $A \otimes 1$
etc. In a basis for $V_{\Lambda_2}$ with the ordering of weight vectors
given by $\{v_{-3},v_{-2},\ldots,v_3 \}$ we have the explicit expressions
\begin{eqnarray*}
\sigma & = & \diag(-\I,1,-\I,1,\I,1,\I), \\
g & = & \diag(q^{-5/2},q^{-2},q^{-1/2},1,q^{1/2},q^2,q^{5/2}),
\end{eqnarray*}
and $F$ is the matrix with $1$ along the anti-diagonal and $0$ elsewhere.
The crossing-relation (\ref{eqn:cross}), which is in a ``standard form''
(apart from the minus sign),
can be obtained from the one given in \cite{Kuniba91}
\begin{displaymath}
R_{12}(u) = - \left( (\sigma g)^{-1} \otimes (\sigma g)\right)
  \stackrel{1}{F} R_{12}^{t_2}(-u-\rho) \stackrel{1}{F}
             \left( (\sigma g) \otimes (\sigma g)^{-1}\right),
\end{displaymath}
by using the symmetry relation $[R(u),(\sigma g) \otimes (\sigma g)]=0$,
which in turn can be checked explicitly.

For an $R$-matrix satisfying properties (\ref{eqn:unit}) to (\ref{eqn:cross})
the Sklyanin transfer matrix
\begin{equation}
t(u,\boldomega) = \tr_a \stackrel{a}{K^+}(u) T_a (u,\boldomega)
      \stackrel{a}{K^-}(u) \tilde{T}_a(u,\boldomega),
\label{eqn:skl}
\end{equation}
with monodromy matrices defined as
\begin{eqnarray}
T_a(u,\boldomega) & = & R_{a1}(u+\omega_1) R_{a2}(u+\omega_2)
   \cdots R_{aN}(u+\omega_N), \nonumber \\
\tilde{T}_a(u,\boldomega) & = &  R_{Na}(u-\omega_N) \cdots
   R_{2a}(u-\omega_2) R_{1a}(u-\omega_1),\label{eqn:monod}
\end{eqnarray}
forms a commuting family $[t(u,\boldomega),t(v,\boldomega)]=0$
if the $K$-matrices $K^{\pm}(u)$ satisfy the boundary YBEs
\cite{Sklyanin88,Mezincescu91a}
\begin{eqnarray}
\lefteqn{
R_{12}(u-v) \stackrel{1}{K^-}(u) R_{21}(u+v) \stackrel{2}{K^-}(v) = }
\hspace{30pt}\nonumber\\
& &
  \stackrel{2}{K^-}(v) R_{12}(u+v) \stackrel{1}{K^-}(u) R_{21}(u-v),
\label{eqn:bybe}\\
\lefteqn{
R_{12}(-u+v) \stackrel{1}{(K^+)^{t_1}}(u) \stackrel{1}{M^{-1}}
R_{21}(-u-v-2\rho) \stackrel{1}{M} \stackrel{2}{(K^+)^{t_2}}(v) = }
\hspace{30pt}\nonumber\\
& &
  \stackrel{2}{(K^+)^{t_2}}(v) \stackrel{1}{M} R_{12}(-u-v-2\rho)
  \stackrel{1}{M^{-1}} \stackrel{1}{(K^+)^{t_1}}(u) R_{21}(-u+v),
  \end{eqnarray}
with $M \equiv - V^t V = M^t$. Only the $K^-(u)$ equation (\ref{eqn:bybe})
needs  consideration since the $K^+(u)$ equation is related to it
by the automorphism $K^+(u) = K^-(-u-\rho)^t M$.

Due to the property (\ref{eqn:comm}) we have immediately the
``trivial'' solution $\{K^-(u)=1, K^+(u) = M\}$ \cite{Mezincescu91b}, which
leads to a $U_q(G_2)$-invariant open spin chain. To obtain the most general
diagonal solution we need to solve the boundary YBE (\ref{eqn:bybe})
with the explicit form of $R(u)$. This leads to a system of 63 coupled
functional equations for the non-zero entries of $K^-(u)$ which we solve
with the help of Mathematica. The simplest two equations couple only the
entry $K^-(u)_{11}$ to $K^-(u)_{22}$ and $K^-(u)_{33}$
respectively. Without loss of generality we set $K^-(u)_{11}=1$.
We are also interested only in solutions with ``initial condition'' $K^-(0)=1$.
These equations can then be easily solved to obtain
\begin{equation}
K^-(u)_{22}= \frac{1+ c_2 q^{2u}}{1+c_2 q^{-2u}},\hspace{10pt}
K^-(u)_{33}= \frac{1+ c_3 q^{2u}}{1+c_3 q^{-2u}},
\end{equation}
where $c_2$ and $c_3$ are arbitrary parameters.
Consideration of the equations which couple $K^-(u)_{22}$ and
$K^-(u)_{33}$ to $K^-(u)_{44}$ leads to
restrictions on the coefficients $c_i$; we find that either $c_2=0$,
$c_2=c_3=0$ or $c_2=c_3=\pm q$. The last two choices are ruled out by
consideration of the equations which couple in $K^-(u)_{55}$. The second
choice eventually leads to $K^-(u)=1$ while the first choice leads to only
two other inequivalent solutions which can be expressed in the form
\begin{equation}
K^-(u) = \Gamma_{\pm}(u) = \diag(1,1,\Psi_{\pm},\Psi_{\pm},
   \Psi_{\pm},q^{4u},q^{4u}),
\end{equation}
with $\Psi_{\pm}=(q \pm q^{2u})(q \pm q^{-2u})^{-1}$. The existence of three
diagonal $K$-matrices for $G^{(1)}_2$ is directly
analogous to the $A_2^{(2)}$ model \cite{Mezincescu91b}, and in contrast to
models like $A^{(1)}_n$ for which there are free parameters in the $K$-matrices
\cite{deVega93}.

\section{Transfer matrix diagonalization}

We consider the diagonalization of the Sklyanin transfer matrix
corresponding to the three cases
\begin{eqnarray*}
{\rm (i)} & \hspace{10pt} &  K^-(u)=1, \; K^+(u)=M,\\
{\rm (ii)} & \hspace{10pt} & K^-(u)=\Gamma_+(u), \;
   K^+(u)=\Gamma_+(-u-\rho)^t M,\\
{\rm (iii)} & \hspace{10pt} & K^-(u)=\Gamma_-(u), \;
   K^+(u)=\Gamma_-(-u-\rho)^t M,
\end{eqnarray*}
i.e.\ boundaries of ``non-mixed'' type.
{}From the properties (\ref{eqn:comm}) to (\ref{eqn:cross}) of the $R$-matrix
can be inferred several properties of the Sklyanin transfer matrix
(\ref{eqn:skl}). Firstly, due to (\ref{eqn:cross}), there is crossing-symmetry
\begin{equation}
t(u,\boldomega) = t(-u-\rho,\boldomega).
\label{eqn:check1}
\end{equation}
The proof is
a simple generalization of that given in \cite{Mezincescu92b} for the
case $K^-(u)=1$. We also have the fusion equation
\begin{eqnarray}
\tilde{t}(u,\boldomega) & = &
   \xi(2u+2\rho) t(u,\boldomega) t(u+\rho,\boldomega) - \nonumber\\
& & \prod_{j=1}^N \xi(u+\rho+\omega_j) \xi(u+\rho-\omega_j) \Delta\{K^-(u)\}
   \Delta\{K^+(u)\},
\label{eqn:check2}
\end{eqnarray}
which relates $t(u,\boldomega)$ to the transfer matrix
$\tilde{t}(u,\boldomega)$ for the fused model \cite{Mezincescu92a}. In equation
(\ref{eqn:check2}) $\Delta\{K^{\pm}(u)\}$ are quantum determinants given by
\begin{eqnarray}
   \Delta\{K^+(u)\} & = & \tr_{12} \left\{ \tilde{{\cal P}}_{12}^-
   \stackrel{1}{V} \stackrel{2}{V} \stackrel{2}{K^+}(u+\rho)
   \stackrel{2}{M^{-1}} R_{12}(-2u-3\rho) \stackrel{2}{M} \stackrel{1}{K^+}(u)
   \right\},\nonumber\\
   \Delta\{K^-(u)\} & = & \tr_{12} \left\{ \tilde{{\cal P}}_{12}^-
   \stackrel{1}{K^-}(u) R_{21}(2u+\rho) \stackrel{2}{K^-}(u+\rho)
   \stackrel{1}{V} \stackrel{2}{V} \right\},
\end{eqnarray}
with the projector $\tilde{{\cal P}}_{12}^- = \frac{1}{7} \stackrel{1}{V}
P_{12}^{t_2} \stackrel{1}{V}$. Define the function $g(u)=[1-u][6-u][4+u]$.
By explicit calculation we find
\begin{eqnarray}
\Delta\{K^-(u)\} & = & \beta_+(u) \; g(2u+\rho) \nonumber\\
\Delta\{K^+(u)\} & = & \beta_-(u) \; g(-2u-3\rho)
\end{eqnarray}
where for each case (in an obvious notation)
\begin{eqnarray*}
\beta_+ (u) & = & \left\{ \begin{array}{l}
 1 \\
 q^{6+4u}(1 \pm q^{7+2u})(1 \pm q^{11+2u})(1 \pm q^{13+2u})
    (1 \pm q^{5+2u})^{-1}
  \end{array}\right.\\
\beta_-(u) & = & \left\{ \begin{array}{l}
 1 \\
 q^{-64-8u} (1 \pm q^{11+2u})(1 \pm q^{13+2u})
    (1 \pm q^{17+2u})(1 \pm q^{19+2u})^{-1}
  \end{array} \right.
\end{eqnarray*}

The two relations (\ref{eqn:check1}) and (\ref{eqn:check2}) provide powerful
constraints on the eigenvalues $\Lambda(u,\boldomega)$ of $t(u,\boldomega)$.
They are key ingredients in the analytic Bethe ansatz for open boundaries
formulated in \cite{Mezincescu92b}. In fact, together with the condition of
periodicity, an analysis of asymptotic behaviour and the ``dressing
hypothesis'', the eigenvalue expression for the Izergin-Korepin model with
$K^-(u)$ can be derived.  As we will soon explain (see also \cite{Artz94}),
this procedure is not adequate in general and has to be supplemented by an
extra assumption.

Define $F(u)=\prod_{j=1}^N[u+\omega_j][u-\omega_j]$ and let
$\Phi_j(u,\boldomega)$ be related to $F(u)$ as $\phi_j(u,\boldomega)$ is
related to $f(u)$ in (\ref{eqn:rel}). Define also
\begin{equation}
D^{(i)}(u) = \prod_{j=1}^{N_i} [u+\I u_j^{(i)}] [u-\I u_j^{(i)}]
\end{equation}
for $i=1,2$,
and let $A_j(u)$ be related to $D^{(i)}(u)$ as $a_j(u)$ is related to
$d^{(i)}(u)$ in (\ref{eqn:dD}). The functions $\Phi_j(u,\boldomega)$ and
$A_j(u)$ are the ``doubled'' versions of $\phi_j(u,\boldomega)$ and $a_j(u)$
respectively. Our ansatz for the eigenvalue $\Lambda(u,\boldomega)$ is
\begin{equation}
\Lambda(u,\boldomega) = \sum_{j=-3}^3 \alpha_j(u) \Phi_j(u,\boldomega) A_j(u),
\label{eqn:eig}
\end{equation}
where $\alpha_j(u)$ are functions independent of lattice size $N$ and
inhomogeneities $\boldomega$. This is the correct form in all known cases,
and can be referred to as the ``doubling hypothesis'' (cf.\ \cite{Artz94}).

In the analytic Bethe ansatz approach introduced in \cite{Mezincescu92b}
$\alpha_j(u)$ and $A_j(u)$ in (\ref{eqn:eig}) are unspecified to begin with.
The functions $\alpha_j(u)$ are determined by calculating the eigenvalue on the
reference state $|\Omega\rangle$, for which $A_j(u)=1$. For a model like the
Izergin-Korepin model where the corresponding $\Phi_j(u,\boldomega)$ are all
{\em distinct} this gives $\alpha_j(u)$ unambiguously \cite{Mezincescu92b}.
The functions $A_j(u)$ can then be determined by using the fusion equation,
crossing-symmetry etc., as described earlier. In our case, and also the case
\cite{Artz94} for $A^{(2)}_{2n}$ for $n>1$,
the $\Phi_j(u,\boldomega)$ are not all
distinct. In particular, since $\Phi_{-2}(u,\boldomega)$ and $\Phi_{-1}
(u,\boldomega)$ are identical, it is possible only to obtain the combination
$\alpha_{-2}(u) + \alpha_{-1}(u)$ unambiguosly. Nevertheless, we can obtain
some of the $\alpha_j(u)$; this we do by explicitly calculating $\langle
\Omega| t(u,\boldomega) |\Omega \rangle$ ($|\Omega \rangle$ being the state
$v_{-3} \otimes \cdots \otimes v_{-3}$) for small $N$ and choosing the
inhomogeneities appropriately to cancel out relevant $\Phi_j(u,\boldomega)$.
For instance, to obtain $\alpha_{-3}(u)$ it is sufficient to have $N=1$ and
$\omega_1=u$ whereas for $\alpha_0(u)$ we choose $N=2$ and $\omega_1=4+u$,
$\omega_2=2+u$.  In this way we find
\begin{eqnarray}
\alpha_{-3}(u) & = & \frac{[2+2u][7+2u][12+2u]}{[1+2u][6+2u][4+2u]}
   \epsilon_{-3}(u),\\
\alpha_{0}(u) & = & \frac{[12+2u][2u]}{[8+2u][4+2u]} \epsilon_{0}(u),
\end{eqnarray}
where
\begin{eqnarray*}
\epsilon_{-3}(u) & = & \left\{ \begin{array}{l}
 1 \\
 \mp q^{-17-2u}(1 \pm q^{1+2u})^2 (1 \pm q^{5+2u})
    (1 \pm q^{7+2u})^{-1}
  \end{array},\right.\\
\epsilon_0 (u) & = & \left\{ \begin{array}{l}
 1 \\
 \mp q^{-23-2u} (1 \pm q^{5+2u})(1 \pm q^{7+2u})
   \end{array}, \right.
\end{eqnarray*}
together with $\alpha_3(u)=\alpha_{-3}(-u-\rho)$.

At this stage we can perform several highly non-trivial checks. They come
from crossing-symmetry (\ref{eqn:check1}),
the fusion equation (\ref{eqn:check2}) -- more specifically,
analyticity of its right-hand side at $u=-\rho$, and
from choosing alternating inhomogeneities \cite{Yung94}, and are, respectively,
\begin{eqnarray}
\alpha_j(u) & = & \alpha_{-j}(-u-\rho), \label{eqn:nt1}\\
\xi(2u+2\rho) \alpha_{3}(u) \alpha_{-3}(u+\rho) & = &  \Delta\{K^-(u)\}
   \Delta \{K^+(u) \},\\
\alpha_{-3}(u) & =&  \frac{l^1_1(u) r^1_1(u)}{R_{11}^{11}(2u)},\label{eqn:nt3}
\end{eqnarray}
with
\begin{displaymath}
r^b_a (u) = K^-(u)_{ab}, \hspace{10pt}
l^b_a(u) = \sum_{cd} R_{da}^{bc}(2u) K^+(u)_{cd}.
\end{displaymath}
These equations (\ref{eqn:nt1}) to (\ref{eqn:nt3}) are indeed satisfied in
all cases.
It now remains to resolve $\alpha_{-2}(u)+\alpha_{-1}(u)$ and
$\alpha_{1}(u)+\alpha_2(u)$ into their parts. This can be done by
imposing analyticity of the eigenvalue expression (\ref{eqn:eig}). Thus we
find that the resulting system of Bethe ansatz equations is
\begin{eqnarray}
\lefteqn{
\delta_1 \prod_{j=1}^N \frac{[\I u_k^{(2)} + \frac{1}{2} + \omega_j]
   [\I u_k^{(2)} + \frac{1}{2} - \omega_j]}
   {[\I u_k^{(2)} - \frac{1}{2} + \omega_j]
   [\I u_k^{(2)} - \frac{1}{2} - \omega_j]} =}\hspace{70pt}\nonumber\\
&  &
\prod_{j=1}^{N_1} \frac{[\I u_k^{(2)} - \I u_j^{(1)}- \frac{3}{2}]
    [\I u_k^{(2)} + \I u_j^{(1)}- \frac{3}{2}]}
    {[\I u_k^{(2)} - \I u_j^{(1)} + \frac{3}{2}]
    [\I u_k^{(2)} + \I u_j^{(1)} + \frac{3}{2}]}\nonumber\\
& & \times  \prod_{j\ne k}^{N_2} \frac{[\I u_k^{(2)} - \I u_j^{(2)}+ 1]
    [\I u_k^{(2)} + \I u_j^{(2)}+ 1]}
    {[\I u_k^{(2)} - \I u_j^{(2)}- 1]
    [\I u_k^{(2)} + \I u_j^{(2)}- 1]},\label{eqn:bae0}\\
\delta_2 & = &
\prod_{j\ne k}^{N_1} \frac{[\I u_k^{(1)} - \I u_j^{(1)}+ 3]
    [\I u_k^{(1)} + \I u_j^{(1)}+ 3]}
   {[\I u_k^{(1)} - \I u_j^{(1)}- 3]
    [\I u_k^{(1)} + \I u_j^{(1)}- 3]}\nonumber\\
& & \times
\prod_{j=1}^{N_2} \frac{[\I u_k^{(1)} - \I u_j^{(2)}- \frac{3}{2}]
    [\I u_k^{(1)} + \I u_j^{(2)}- \frac{3}{2}]}
    {[\I u_k^{(1)} - \I u_j^{(2)}+ \frac{3}{2}]
    [\I u_k^{(1)} + \I u_j^{(2)}+ \frac{3}{2}]},\label{eqn:bae}
\end{eqnarray}
where the functions $\delta_i$ take different forms in terms of $\alpha_j$,
depending on the specific point at which we consider analyticity. For instance,
analyticity at $u=\I u_k^{(2)}-\frac{1}{2}$ and at $u=\I u_k^{(2)}-\frac{7}{2}$
give rise to, respectively,
\begin{equation}
\delta_1 = \frac{\alpha_{-3}(\I u_k^{(2)}-\frac{1}{2}) [2 \I u_k^{(2)} -1]}
    {\alpha_{-2}(\I u_k^{(2)}-\frac{1}{2}) [2 \I u_k^{(2)} +1]}
        = \frac{\alpha_{-1}(\I u_k^{(2)}-\frac{7}{2}) [2 \I u_k^{(2)} -1]}
    {\alpha_{0}(\I u_k^{(2)}-\frac{7}{2}) [2 \I u_k^{(2)} +1]}
\end{equation}
whereas analyticity at $u=\I u_k^{(1)}-2$ and at $u=-\I u_k^{(1)}-2$ gives
\begin{equation}
\delta_2 = \frac{\alpha_{-2}(\I u_k^{(1)}-2) [2 \I u_k^{(1)} -3]}
    {\alpha_{-1}(\I u_k^{(1)}-2) [2 \I u_k^{(1)} +3]}
        = \frac{\alpha_{-1}(-\I u_k^{(1)}-2) [2 \I u_k^{(1)} -3]}
    {\alpha_{-2}(-\I u_k^{(1)}-2) [2 \I u_k^{(1)} +3]}.
\end{equation}
Such equalities allow $\alpha_{\pm 1}(u)$ and $\alpha_{\pm 2}(u)$ to be
related to $\alpha_{\pm 3}(u)$ and $\alpha_0(u)$ which we have already
determined. Solving the resulting functional equations we find that
\begin{eqnarray}
  \alpha_{-2}(u) & = & \frac{[12+2u][2u][7+2u]}{[1+2u][4+2u][6+2u]}
   \epsilon_{-2}(u), \\
  \alpha_{-1}(u) & = & \frac{[12+2u][2u]}{[4+2u][6+2u]} \epsilon_{-1}(u),
\end{eqnarray}
with $\epsilon_{-2}(u)=\epsilon_{-3}(u)$ and $\epsilon_{-1}(u)=
\epsilon_{0}(u)$,
together with $\alpha_{1}(u)=\alpha_{-1}(-u-\rho)$ and
 $\alpha_{2}(u)=\alpha_{-2}(-u-\rho)$. This completes the determination of
the eigenvalue expression $\Lambda(u,\boldomega)$. We have shown that it
satisfies all the checks mentioned in this paper.  The corresponding
results for the ``boundary factors'' $\delta_i$ in the Bethe ansatz equations
(\ref{eqn:bae0}) and ((\ref{eqn:bae}) are found to be $\delta_1=1$ in all
cases, while $\delta_2=1$ for case (i) and
\begin{equation}
\delta_2  =
    \left( \frac{ q^{\I u^{(1)}_k -3/2} \pm q^{-\I u^{(1)}_k +3/2}}
                { q^{\I u^{(1)}_k +3/2} \pm q^{-\I u^{(1)}_k -3/2}} \right)^2
\end{equation}
in the remaining two cases.
We note that there is a striking resemblance to the
corresponding ``boundary factors'' for the Izergin-Korepin model \cite{Yung94}.

\section{Discussion}

We have seen how the general considerations of \cite{Sklyanin88,Mezincescu91a}
for obtaining integrable models with open boundaries can in principle be
applied to $R$-matrices based on exceptional affine algebras. In
particular we have obtained all the diagonal $K$-matrices for the model based
on $G^{(1)}_2$. With $K^-(u)=1$ and $K^+(u)=M$, the corresponding spin chain
has Hamiltonian $\left. H=\sum_{k=1}^{N-1} \check{R}_{k,k+1}'(u)\right|_{u=0}$
\cite{Mezincescu91c} and is $U_q(G_2)$-invariant. In the rational limit
$q\rightarrow 1$ the Hamiltonian is both $G_2$- and $su(2)$-invariant
\cite{Kennedy92,Batchelor94a} with its energy spectrum determined by the
Bethe ansatz equations (\ref{eqn:bae0}) and (\ref{eqn:bae}) with
$[x]\rightarrow x$, $\delta_i=1$ and $\omega_i=0$.

The method we have used to diagonalize the transfer matrices can be considered
a variation of the analytic Bethe ansatz \cite{Mezincescu92b}, with an
extra (unproven) assumption, namely the ``doubling hypothesis''. A rigorous
alternative method is probably the (nested) Bethe ansatz along the lines of
\cite{deVega94a} which we expect to be much more complicated to apply here.
We have seen how the Bethe ansatz equations for the $G^{(1)}_2$ model
-- in particular, the ``boundary factors'' -- resemble those for $A_2^{(2)}$.
It would be interesting to see if this can be explained on Lie algebraic
grounds alone, analogous to the periodic boundary situation
\cite{Reshetikhin87}.

\noindent
{\bf Acknowledgements}

We are grateful to A. Kuniba and J. Suzuki for helpful correspondence. This
work is supported by the Australian Research Council.

\end{document}